\documentstyle[twoside,epsf]{article}

\input{ibvs2.sty}

\begin{document}

\IBVShead{xxxx}{xx xxx 2003}

\IBVStitle{New R CrB-Type Star H\lowercase{ad}V98}

\IBVSauth{Kato,~Taichi$^1$; Katsumi, Haseda$^2$}
\vskip 5mm

\IBVSinst{Dept. of Astronomy, Kyoto University, Kyoto 606-8502, Japan,
          e-mail: tkato@kusastro.kyoto-u.ac.jp}

\IBVSinst{Variable Star Observers League in Japan (VSOLJ), 2-7-10 Fujimidai,
       Toyohashi City, Aichi 441-8135, Japan,
       e-mail: khaseda@mx1.tees.ne.jp}

\IBVSobj{HadV98}
\IBVStyp{RCB}
\IBVSkey{R CrB star, photometry}

\begintext

   HadV98 = CoD $-$22$^\circ$12017 = GSC 6825.253 is a variable star
discovered by K. Haseda (Haseda and Kato 2001).  The variable had been
almost constant until 2001 April, when it suddenly started fading.
The object was suspected to be an R CrB-type star, from the time
of the discovery alert, based on its light behavior and its Tycho-2
moderate color ($B-V \sim$1.0).
This star has been very recently confirmed to be a genuine R CrB-type
variable from spectroscopy (Hasselbach et al. 2002).

   We studied the variability of this object using Haseda's discovery
material combined with the ASAS-3 survey data (Pojmanski 2002).
The light curve constructed from Haseda's observations and the ASAS-3
public data is presented in Figure 1.  Haseda's observations were
performed with a $D$ = 10 cm f/4.0 telephoto lens and unfiltered
T-Max 400 emulsions.  The passband of observations covers the range of
400--650 nm.  The photographic magnitudes were estimated using GSC
stars.  A 0.08 mag systematic correction was added to Haseda's
photographic observations, in order to best match the ASAS-3 $V$
magnitudes.  The recorded range of variability in $V$ was 10.7--13.4,
but the true minimum seems to have been fainter.

\IBVSfig{10cm}{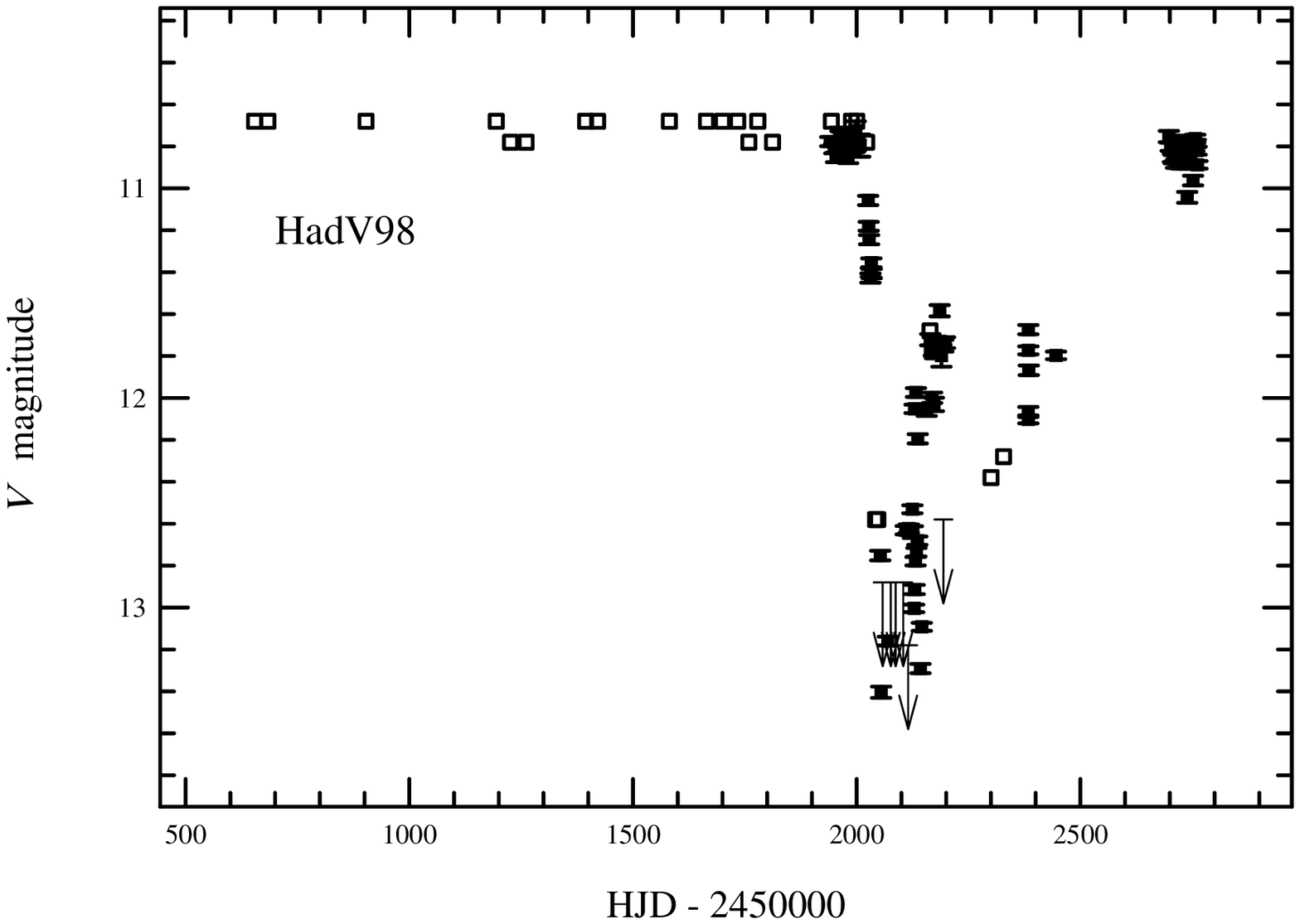}{Light curve of HadV98 drawn from Haseda's
photographic observations and the ASAS-3 $V$-band public data.
The filled squares with error bars and open squares represent
ASAS-3 data and Haseda's measurements, respectively.
The arrows represent upper limit observations by Haseda.
}

   The light curve shows a rapid decline and a slower recovery, which
are characteristic to R CrB-type optical light curves (Clayton 1996).
The 2001--2002 fading was composed of at least two distinct minima
(Figure 2), one lasting in 2001 May -- 2001 July (JD 2452030--2452130)
and another minimum suddenly starting from 2001 October (JD 2452190--),
whose entire picture was not well observed because of the solar
conjunction.
Such an occurrence of closely spaced two minima are also sometimes
seen in other R CrB-type fadings and other dust-forming objects
(Kato et al. 2002).

\IBVSfig{10cm}{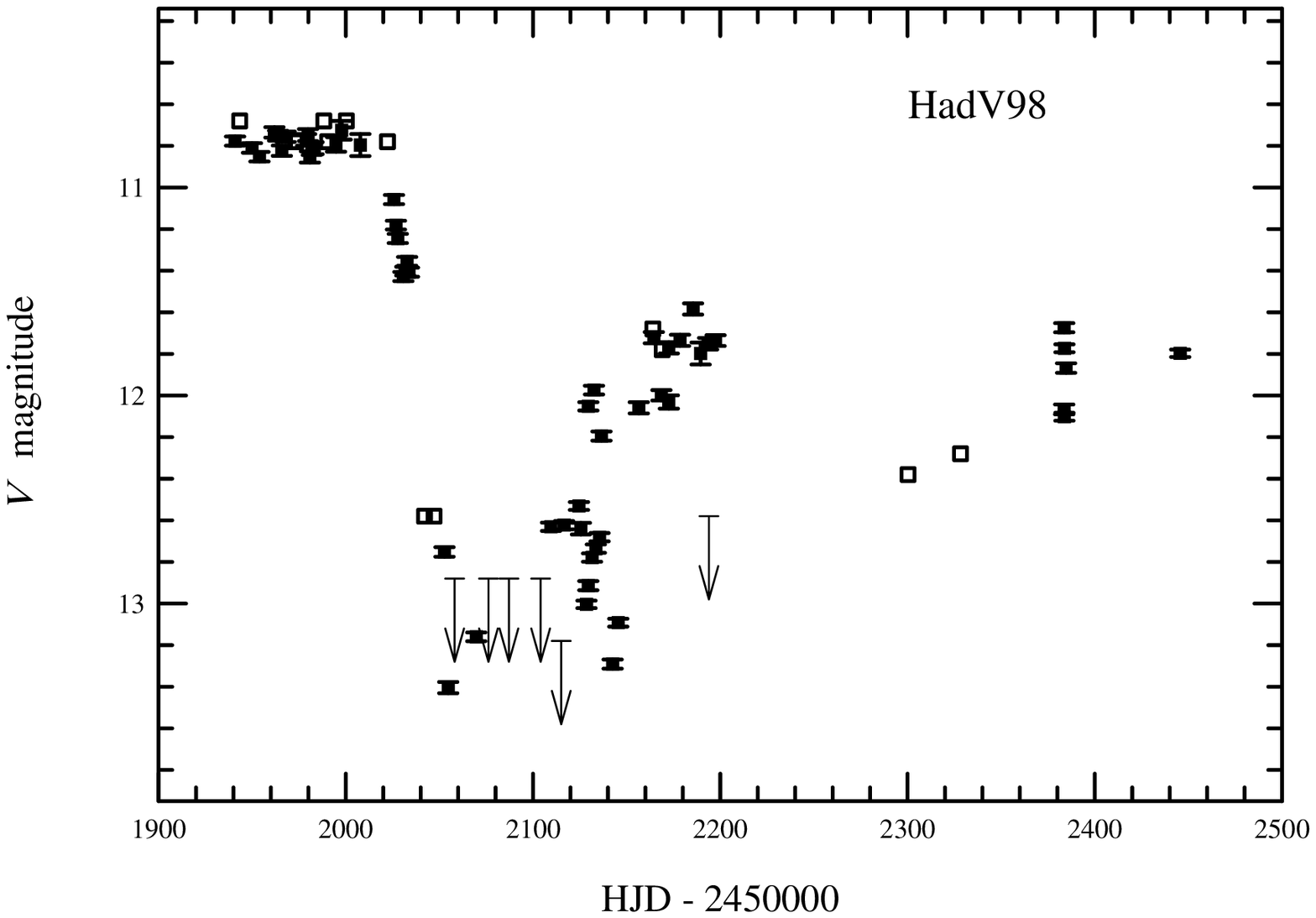}{Enlargement of the fading starting in
2001 May.  The symbols are the same as in Figure 1.
}

   We also searched for a pulsation signal, which is frequently seen
in many R CrB stars, most notably recorded in RY Sgr and V854 Cen
(cf. Clayton 1996), using the ASAS-3 observations at maximum between
JD 2452697 and 2452762 (20 measurements).  No remarkable pulsation
was found with an amplitude larger than 0.1 mag.  This finding is
consistent with the lack of remarkable variability in Haseda's
observations before 2001.  HadV98 apparently belongs to a group
of R CrB stars showing less distinct pulsations at maximum.

\vskip 5mm

We are grateful to G. Pojmanski for making the ASAS-3 survey data
publicly available, and generously allowing us for unlimited usage.
This work is partly supported by a grant-in-aid (13640239, 15037205)
from the Japanese Ministry of Education, Culture, Sports, Science and
Technology.

\references

Clayton, G. C., 1996, PASP, 108, 225

Haseda, K., Kato, T., 2001, vsnet-alert 6226 \\
  (http://www.kusastro.kyoto-u.ac.jp/vsnet/Mail/alert6000/msg00226.html)

Hasselbach, E., Clayton, G. C., Smith, P. S, 2002, AAS Meeting 201, 17.11

Kato, T., Haseda, K., Takamizawa, K., Yamaoka, H., 2002, A\&A, 393, 69L

Pojmanski, G. 2002, Acta Astronomica, 52, 397

\endreferences

\end{document}